\documentclass[review]{elsarticle}

\makeatletter
\def\ps@pprintTitle{%
 \def\@oddfoot{}%
 \let \@evenfoot\@oddfoot}
\makeatother

\usepackage{lineno,hyperref}
\usepackage{lineno,hyperref}

\usepackage{amsmath,amssymb}

\usepackage{tabularx}
\usepackage[ruled]{algorithm2e} 
\usepackage{algorithmic}

\usepackage{mathtools}
\usepackage{amsmath} 
\usepackage{amsthm}

\usepackage{amsfonts}
\usepackage{tikz}
\usepackage[utf8]{inputenc}

\newtheorem{definition}{Definition} [section]

\newtheorem{theorem}{Theorem} [section]
\newtheorem{corollary}{Corollary} [section]
\newtheorem{lemma}{Lemma} [section]

\theoremstyle{definition}

\usepackage{booktabs}

\begin{document}

\begin{frontmatter}

\title{A note on Finiteness and Boundedness in Cartesian Products of Discrete and Well-Ordered Sets, Application to Transition Systems}
\author{Gérard Memmi}
\address{LTCI, Telecom-Paris, Institut polytechnique de Paris}
\date{\today}
\begin{abstract}
\textit{In this short note, we prove that any closed interval in a well-ordered and discrete space is finite.
The coincidence between finiteness and boundedness can then be established for transition systems where the state space is well-ordered and discrete.
Finally, this result is also proven when the state space is a Cartesian product of $d$ discrete and well-ordered sets.}
\end{abstract}

\begin{keyword}
Finiteness, Boundedness, Transition System, well-ordered set, discrete space, compact. 
\end{keyword}

\end{frontmatter}

\tableofcontents


\section{Introduction}
This short note establishes the coincidence between finiteness and boundedness first, for transition systems where the state space is well-ordered and discrete, then this result is also proven when the state space is a Cartesian product of a finite number of discrete and well-ordered sets.

These results were motivated by the \textit{boundedness problem} for well-structured transition systems (WSTS) as specified in \cite{FinkelS01,BFS24}. 
To decide whether a transition system $M$ and an initial state $q_{init}$ are such that the set of reachable states from $q_{init}$ (denoted by $RS(M,q_{init})$) is finite, is defined as the boundedness problem. 
The problem statement and its terminology is inspired by Petri \cite{BR82} nets or Vector Addition Systems (VAS) \cite{KM69} theory; it suggests a kind of equivalence between finiteness and boundedness, which is true in the context of Petri nets, but generally false. 
This problem first resolved by R. Karp and R. Miller (see \cite{KM69}) for VAS, has then been resolved for a large number of classes of WSTS (see \cite{FinkelS01}), itself a class of transition systems described section \ref{subsubsec: WSTS}. 
The constraints of discreteness and well-ordering especially with respect to the factors of a finite Cartesian product, are not unreasonable in the context of transition systems or discrete event systems (DES) \cite{CL10} where a large number of sub-models (or classes) including Petri nets verify them (for instance, adding discrete parameters to Petri nets as in \cite{David15}).
This motivated the introduction of State Variable Transition Systems, section \ref{subsec: SVTS} in this note and served as the starting point for this note.
 
In section \ref{sec: background}, we recall the definition of various order relations and elements of topology that we will need to prove our main results. We also recall section \ref{subsec: TS&WSTS} the definitions of transition systems and well-structured transition systems, then introduce section \ref{subsec: SVTS}, transition with state variables (SVTS) as a generalization of fair transition systems. 
In section \ref{sec: Main results}, we start with theoretical results leading to the conditions under which a closed interval collapses to a finite set, establishing that the combination of compactness and discreteness grants finiteness.
Then, section \ref{subsec: finiteness and boundedness}, we apply our result to transition systems \cite{K76}, introducing a new transition system directly derived from Fair Transition Systems \cite{MannaP93} and conclude with potential future works section \ref{sec: conclusion}

\section{Background}
\label{sec: background}
Basic notions of orderings and topology can be found in \cite{Munkres02,W04}.
More definitions and properties on ordered sets can be found in \cite{KRUSKAL72}.
We consider a set $\emph{Q}$ equipped with two binary relations denoted by $\leq$ and $\rightarrow$. 
We limit ourself to recall definitions of ordered sets using the notation $\leq$ that will be used for transition systems later in this note. 
We conclude this section with the notion of transition systems using the notation $\rightarrow$.

\subsection{Definitions and properties of various ordered sets}
\label{subsec: def of Ordered sets}
\begin{definition}[Quasi-Ordered Sets and Partially Ordered Sets]
\label{def: qoset&poset}
A \textit{quasi-ordered set} (qoset) is a set $\emph{Q}$ equipped with a reflexive and transitive relation denoted by $\leq$.

A \textit{partially-ordered set} (poset) $\langle \emph{Q}, \leq \rangle$ is a qoset such that $\leq$ is antisymmetric.
\end{definition}
\begin{definition} [Totally Ordered Sets]
\label{def: totally-ordered}
$\langle \emph{Q}, \leq \rangle$ is a \textit{totally ordered set} is a poset such that every pair of elements of $\emph{Q}$ is comparable with $\leq$.

A closed interval $[a, b]$ in $\langle \emph{Q}, \leq \rangle$ is a set $\{x \in \emph{Q}\ |\ a \leq x \leq b\}$.
An open interval $(a,b)$ in $\langle \emph{Q}, \leq \rangle$ is then a set $[a, b]\setminus \{a,b\}$.
\end{definition}
A totally ordered set is sometimes called linearly ordered or simply ordered as in \cite{bha19,Munkres02}. A totally ordered set may have the least upper bound property (LUB) (i.e. a supremum):
\begin{definition} [Least Upper Bound property (LUB)]
\label{def: LUB}
A totally ordered set $\langle \emph{Q}, \leq \rangle$ has the LUB property if any non-empty subset of $\emph{Q}$ that is bounded above, has a least upper bound:
if $\forall A \subseteq \emph{Q}\setminus \emptyset,$ such that $\exists s \in \emph{Q},\ \forall a \in A,\  a \leq s$, then $\exists u \in \emph{Q}$, such that $\forall s' \in \emph{Q},\ \forall a \in A,\  a \leq s'$, we have $u \leq s'$.
\end{definition}
\begin{definition}[Well-Ordered Set]
\label{def: WO}
$\langle \emph{Q}, \leq \rangle$ is a \textit{well-ordered set} if is a totally ordered set and every non-empty subset of $\emph{Q}$ contains a unique least element.
\end{definition}

\begin{definition}[Well-Quasi-Ordered Set]
\label{def: WQO}
$\langle \emph{Q}, \leq \rangle$ is a \textit{well-quasi-ordered set} if
 is a qoset and for any infinite sequence $q_1,...q_k,...$ there exist two indices i, j such that $i < j$ and $q_i \leq q_j$.
\end{definition}

Equivalently, a well-quasi-ordered set is such that every strictly descending sequence is finite and every antichain is finite. Other equivalent definitions can be found in \cite{Milner85}.
\begin{lemma} [well-quasi-order is preserved under finite Cartesian product]
\label{lem: WQO in Cartesian product}
The finite Cartesian product of well-quasi ordered sets is well-quasi ordered, consequently, since well-ordered sets are well-quasi ordered, the finite Cartesian product of well-ordered sets is well-founded. 
\end{lemma}
This result is mentioned in \cite{KRUSKAL72} and does not hold if the Cartesian product is infinite, since the set would have obvious infinite antichains.

\begin{definition}[Well-founded Set]
\label{def: WFO}
$\langle \emph{Q}, \leq \rangle$ is a \textit{well-founded set} if every non-empty subset of \emph{Q} has at least one minimal element.
\end{definition}
Equivalently, $\langle \emph{Q}, \leq \rangle$ is well-founded if it does not admit an infinite strictly descending sequence.
Note that this definition of a well-founded set does not specify any property (such as reflexivity,..),for the relation $\leq$ and if $A \subseteq \emph{Q}$, $x$ is a minimal element of $A$ means $\{q \in B\setminus\{x\} \ |\ q \leq x\}=\emptyset$. 

If a set is well-ordered then it is well-quasi ordered and if it is well-quasi ordered then it is well-founded. The reverse does not hold. 

Just as in lemma \ref{lem: WQO in Cartesian product}, the finite Cartesian product of well-quasi ordered sets is well-founded, consequently, since well-ordered sets are well-quasi ordered, the finite Cartesian product of well-ordered sets is well-founded under the componentwise order. 
This does not hold if the Cartesian product is infinite. Counterexamples exist although the argument on the infinite antichain does not work for well-founded sets.

\subsection{Basic elements of topology}
\label{subsec: Topology}
\begin{definition}[Order Topology]
\label{def: order-topology}
The order topology on a totally ordered set $\langle \emph{Q}, \leq \rangle$ is the topology generated by the subbasis consisting of open rays of the form $(x, \infty)$ and $(-\infty, x)$.

\end{definition}
\begin{definition}[Compactness]
\label{def: compactness}
A subset $A$ of $\emph{Q}$ is a compact set if any open cover of $A$ has a finite subcover of $A$.
\end{definition}

We can find the following theorem in \cite{Munkres02} p.172 or in \cite{bha19} with a shorter proof.
\begin{theorem}[Compactness of Intervals in totally ordered sets]
\label{th: Compactness+LUB}
Let $\langle Q, \le \rangle$ be a totally-ordered set having the LUB property,  equipped with the order topology. Then any closed interval $[a, b] \subseteq \emph{Q}$ is compact.
\end{theorem}
We can find the following theorem in \cite{Munkres02} p.234.
\begin{theorem}[Tychonoff]
\label{th: Tychonoff}
An arbitrary product of compact spaces is compact in the product topology.
\end{theorem}
\begin{definition}[Discrete Space]
\label{def: discreteness}
A topological space $\emph{Q}$ is discrete if every singleton subset $\{x\}$ of $\emph{Q}$ is an open. 
\end{definition}
Equivalently, the collection of all subsets of $\emph{Q}$ is a discrete topology \cite{Munkres02}. 
\begin{theorem} [Discreteness and Cartesian product]
\label{th: Discrete in Cartesian product} 
A Cartesian product of $d$ discrete spaces is discrete (under the condition that $d$ is finite).
 
If a non-empty Cartesian product is a discrete space then its factors are also discrete. 
\end{theorem}
The proof is easy and comes from the facts that a tuple $x = (x_1,...x_d)$ can be seen as a Cartesian product of singletons: $x = \prod_{i=1} ^{i=d}\{x_i\}$ and that a projection map is an open map (see \cite{Munkres02} chapter 2 paragraph 19 on product topology)
\hfill $\square$
\subsection{Transition Systems (TS) and Well-Structured Transition Systems (WSTS)}
\label{subsec: TS&WSTS}
In order to model and analyze programs and concurrent systems, many computer science papers utilize the notion of Transition Systems sometimes more appropriately called State Transition Systems as in \cite{ChMi88}, sometimes without even providing a formal definition.
In what follows, we first use the general definition introduced in \cite{Kel72,K76}. 
We then present two classes of Transition Systems sections \ref{subsubsec: WSTS} and \ref{subsec: SVTS}. 
The later will serve as the basis for the final section of this note, which is devoted to our contribution to boundedness and finiteness.
 
\subsubsection{Transition Systems (TS)}
\label{subsubsec: transition systems}

\begin{definition} [Transition Systems]
\label{def: TS}
A \textit{Transition System} is a pair $M=\langle\emph{Q}, \rightarrow \rangle$ where: 

$\emph{Q}$ is a non empty \textit{state space}, a state is an element of $\emph{Q}$,

$\rightarrow$ is a binary relation over $\emph{Q}$ and is called the \textit{set of transitions}.

A pair of states $(q,q')$ such that $(q,q') \in \rightarrow $ is denoted by $q \rightarrow q'$.
$ \overset{*}{\rightarrow}$ denotes the reflexive and transitive closure of ${\rightarrow}$ and we say that a state $q'$ is reachable from the state $q$ if and only if $q \overset{*}{\rightarrow} q'$.
\end{definition}
Note that $\langle \emph{Q}, \overset{*}{\rightarrow} \rangle$ is a qoset by definition and $\emph{Q}$ as well as $\rightarrow$ can be infinite.

If $Init$ is a non-empty subset of $\emph{Q}$, then $RS(M,Init)$ denotes the set of all reachable states from $Init$: $RS(M,Init) = \{q \in \emph{Q} \ |\ \exists q_{init} \in Init, \ q_{init} \overset{*}{\rightarrow} q\}$. Note that since $\overset{*}{\rightarrow}$ is reflexive, $Init \subseteq RS(M,Init)$, hence $RS(M, Init)$ is non-empty.
 
At this stage, neither states nor transitions are precisely defined. This is a deliberate choice intended to make Transition Systems as general as possible. 
It is therefore not surprising that a number of behavioral properties problems are not decidable for $TS$. 
Authors will then expand TS by adding various pieces of information, either by specifying how the set of transitions works as in \cite{FinkelS01} or by refining the state space, especially through the notion of state variables as in \cite{MannaP93,Sankar03}.

\subsubsection{Well Structured Transition Systems (WSTS)}
\label{subsubsec: WSTS}

The concept of well-structured transition systems as a subclass of TS, was mainly motivated by extending the domain of conceptual machines for which boundedness and other behavioral properties would be decidable, (see \cite{FinkelS01} in particular). 
In this regard, the authors enriched a TS with an order relation denoted $\leq$ in $\emph{Q} \times \emph{Q}$. Here we give the definition with the TS notation of the former section:

\begin{definition} [Well Structured Transition Systems (WSTS)]
A \textit{well-structured transition system} (WSTS) is a transition system $M = \langle \emph{Q},\rightarrow,\leq  \rangle$  where:

$ \langle  Q,\rightarrow  \rangle$ is a TS, 

$\langle \emph{Q}, \leq \rangle$ is a well-quasi ordered set,

$\forall (q,q',s) \in \emph{Q}^{3}$ such that $q \leq q'$ and $q \rightarrow s$ then $\exists s' \in \emph{Q}$ such that $q' \overset{*}{\rightarrow} s'$ and  $s \leq s' $. 
\end{definition}
The third item of this definition suggests a form of monotonicity for $\rightarrow$.

Interesting classes of WSTS were recently introduced, namely: 
trace bounded $\infty$-effective cd-WSTS \cite{CFS16} and very-WSTS \cite{blfigo18}
for which the covering graph can be constructed.

\section{Main Results}
\label{sec: Main results}
Our main results can be seen in two parts: first, by following-up section \ref{subsec: def of Ordered sets} and \ref{subsec: Topology}; second, by applying these results to transition systems described section \ref{subsec: TS&WSTS}.
\subsection{Fundamental results}
\label{subsec: Fundamentals}
\begin{lemma}
\label{lem: WO has LUB} 
The LUB property holds for any well-ordered set $\langle Q, \leq \rangle$.
\end{lemma}
For any non-empty set $A$ of $\emph{Q}$, we can associate the subset $B$ of its upper bounds. If $B$ is non-empty (i.e. if $A$ is bounded above) then $B$ has a unique least element by definition \ref{def: WO}
\hfill$\square$
\begin{corollary}
\label{cor: Compactness+WO}
Let $\langle Q, \leq \rangle$ be a well-ordered set equipped with the order topology. Then any closed interval $[a, b] \subseteq \emph{Q}$ is compact.
\end{corollary}
This result is directly derived from the fact that a well-ordered set has the LUB property by lemma \ref{lem: WO has LUB}, then by applying theorem \ref{th: Compactness+LUB}
\hfill$\square$



\begin{theorem}[Compactness and Discreteness Implies Finiteness]
\label{th: Compact+discrete = finite}

Let $[a, b]$ be any topological space. If $[a, b]$ is both compact and discrete, then $[a, b]$ is a finite set.
\end{theorem}

Let $[a, b]$ be a discrete topological space. By definition \ref{def: discreteness}, for every element $x \in [a, b]$, the singleton set $\{x\}$ is an open set. 
We construct a collection $\mathcal{C}$ consisting of all such singletons:
\[ \mathcal{C} = \{ \{x\} : x \in [a, b] \} \]
Since $\bigcup_{x \in [a, b]} \{x\} = [a, b]$, the collection $\mathcal{U}$ forms an open cover of $[a, b]$.

$[a, b]$ is compact, by definition \ref{def: compactness}, every open cover must admit a finite subcover.
Being finite composed of singletons and covering $K$ proves that $K$ is finite
\hfill $\square$

From corollary \ref{cor: Compactness+WO} and theorem \ref{th: Compact+discrete = finite}, we can easily deduce the following corollary:
\begin{corollary} [well-ordered and discreteness implies finiteness]
\label{cor: WO+Discrete is finite}
If $\langle \emph{Q}, \leq \rangle$ is a well-ordered and discrete space, then any closed interval $[a,b]$ is finite.
\end{corollary}
From corollary \ref{cor: Compactness+WO}, we have $[a,b]$ compact and can apply theorem \ref{th: Compact+discrete = finite} to deduce that $[a,b]$ is finite
\hfill $\square$

\subsection{On Finiteness and Boundedness in Transition Systems}
\label{subsec: finiteness and boundedness}

We first define finiteness and boundedness within the general framework of transition systems as defined section \ref{subsubsec: transition systems}. 
We then present two settings for transition systems and SVTSs in which finiteness and boundedness coincide with theorems \ref{th: Boundedness and Finiteness in TS} \ref{th: finiteness=boundedness in SVTS}. 

The concepts of finiteness and boundedness for transition systems are defined with respect to a set of reachable states from either a subset of states (namely, the initial subset called $init$) or more rarely, from the entire state state space $\emph{Q}$. 
In this last case, we speak of structural finiteness or structural boundedness.
Interestingly, for a number of classes of transition systems including Petri nets, structural boundedness is computable (see \cite{BR82,memmi24}).
\begin{definition} [Finiteness in Transition systems]
\label{def: finiteness of TSs}
Given a transition system $M= \langle \emph{Q}, \rightarrow \rangle$ and $Init$ a subset of $\emph{Q}$, we say that 
$M$ is Init-finite (or finite) if and only if $\ \forall q_{init} \in Init, \ RS(M, q_{init})$ is finite.

$M$ is structurally finite if and only if $\forall q \in \emph{Q}, M$ is \{$q$\}-finite. 
\end{definition}
Note that $RS(M,Init)$ can be infinite and $M$ be Init-finite.

\begin{definition} [Boundedness in Transition systems]
\label{def: Boundedness of TSs}
Given a transition system $M= \langle \emph{Q}, \rightarrow \rangle$, a relation $\leq$, and $Init$ a subset of $\emph{Q}$, we say that 
$M$ is Init-bounded (or bounded) if and only if $\ \forall q_{init} \in Init, RS(M, q_{init})$ is bounded above.

$M$ is structurally bounded if and only if $\forall q \in \emph{Q}, M$ is \{$q$\}-bounded. 
\end{definition}
Strictly speaking, we should say that $M$ is \textit{bounded above}. 
However, this terminology comes from Petri net theory, and we will see that theorem \ref{th: finiteness=boundedness in SVTS} justifies this misuse of term.
\begin{theorem} [Boundedness and Finiteness in TS]
\label{th: Boundedness and Finiteness in TS}
    In a transition system $M =\langle \emph{Q}, \rightarrow \rangle$ where $\emph{Q}$ is equipped with a relation $\leq$ such that $\langle \emph{Q}, \leq \rangle$ is a well-ordered and discrete space,
    we have:

    $M$ is (structurally) bounded if and only if $M$ is (structurally) finite.
\end{theorem}
The same proof can be used for proving that finiteness and boundedness coincide or that structural finiteness and structural boundedness coincide.

As a well-ordered set, $\langle \emph{Q}, \leq \rangle$ has a unique least element $l \in \emph{Q}$. If $M$ is bounded, then when have:
$\forall q \in Init$, 
$\exists r \in \emph{Q}$ and $ \ RS(M,q) \subseteq [l,r]$. Since $\langle \emph{Q}, \leq \rangle$ is well-ordered and discrete, we can apply corollary \ref{cor: WO+Discrete is finite} to deduce that $[l,r]$ is finite; and since $ \ RS(M,q) \subseteq [l,r]$, $M$ is finite.

Reversely, if $M$ is finite and $\leq$ being a total order, every pair of elements of $RS(M, Init)$ are comparable then $RS(M, Init)$ has a greatest element. This holds for all finite subset including $\emph{Q}$ itself
\hfill
$\square$

However, we cannot conclude if instead of well-ordered, $\langle \emph{Q}, \leq \rangle$ was well-quasi ordered (as in WSTS) since a pair of states could be incomparable. An additional structure of a kind of lattice (for instance, if $\forall q_1, q_2 \in \emph{Q}, \ \exists q_3 \in \emph{Q}, \ q_1 \leq q_3$ and $q_2 \leq q_3$) would then be necessary to prove that finiteness implies boundedness.

\subsection{Finiteness and Boundedness in State variable Transition Systems  (SVTS)}
\label{subsec: SVTS}
A state variable can represent any kind of information even outside of the modeled system itself (such as an ambient temperature) that is considered useful in regard to the understanding of the behavior of the modeled system.
State variables vary within a domain of values and help redefining the concept of state. 
This important addition to classic transition systems will allow one to speak of assertions or invariants as logical, temporal, or algebraic expressions over a set of state variables (see, for instance, \cite{MannaP93,memmi24}).

Here, we define the concept of State Variable Transition System (SVTS) clearly drawn from Fair Transition Systems introduced by Z. Manna and A. Pnueli \cite{MannaP93}.
SVTS are slightly more general than Fair transition systems, by relaxing the notion of transition to stick to the definition of a set of transitions as in definition \ref{def: TS}) and by denoting the state space by $\emph{Q}$ and defining $Init$ as a subset of the state space (rather than by an ``initial condition``). 
Our result on finiteness and boundedness presented at the end of this section will apply in this setting.

\begin{definition} [State Variables Transition Systems (SVTS)]
\label{def: Q from SV}
A State Variable Transition System (SVTS) is a transition system augmented by a set of state variables $M=\langle \mathcal{SV}, \emph{Q}, \rightarrow \rangle$ where:

$\mathcal{SV} = \{v_0, \cdots, v_i, \cdots\}$ is a set of state variables. An element $v_i$ of $\mathcal{SV}$ is a state variable and varies in a set $D_i$ called the domain of $v_i$,

the state space $\emph{Q}$ of $M$ is defined as the Cartesian product of the domains $D_i$: 

$\emph{Q}= \prod_{v_i \in \mathcal{SV}} D_i$,

a \textit{state} $q$ as an element of the state space $\emph{Q}$ is a tuple $q=(q_0, \cdots , q_i, \cdots)$ and can also be seen as a mapping from $\mathcal{SV}$ to the set of potential values taken by the elements of $\mathcal{SV}$ \footnote{In \cite{MannaP93}, $q$ is a type consistent interpretation of $\mathcal{SV}$}:

$\forall i \in \mathbb{N},\quad q_i=q(v_i)$, and $q(v_i) \in D_i$

$\rightarrow$ is a binary relation over $\emph{Q}$ and is called the \textit{set of transitions}.
\end{definition}

The notion of places in Petri nets theory (see \cite{BR82} for definitions and theoretical results with their proofs), can be seen as state variables, and a Petri nets are  a specific class of SVTS as well as a specific class of WSTS (as mentioned in \cite{FinkelS01}).

For SVTSs equipped with a set of state variables $\mathcal{SV}$, the definitions of boundedness and finiteness of section \ref{subsec: finiteness and boundedness} indeed hold, and assuming that $\mathcal{SV}$ is finite and state variable domains are discrete and well-ordered, theorem \ref{th: Boundedness and Finiteness in TS} can be replace by the following theorem:







\begin{theorem} [Boundedness and Finiteness in SVTS]
\label{th: finiteness=boundedness in SVTS}
    In an SVTS $M=\langle \mathcal{SV}, \emph{Q}, \rightarrow \rangle$ where:
    
    $|\mathcal{SV}| = d$ is finite,
    
    $\emph{Q}$ is a Cartesian product of $d$ discrete and well-ordered sets $D_i$
    we have:

    $M$ is (structurally) bounded if and only if $M$ is (structurally) finite.
\end{theorem}
If $RS(M, Init)$ is finite, then each state variable $v_i$ varies in $X_i$, a finite subset of $D_i$.
$D_i$ is well-ordered and $X_i$ is finite, then $X_i$ is bounded by  $[l_i,r_i] \subseteq D_i$: we can construct $r = (r_1,...r_d)$ componentwise; $r \in \emph{Q}$ since $\emph{Q}= \prod_{i=1}^{i=d} D_i$ and, we have $\forall q \in RS(M, Init), \  q \leq r$.

Reversely, let's consider $RS(M, Init)$ bounded above 
(i.e. $\exists r \in \emph{Q}, \forall q  \in RS(M,Init), \ q \leq r$).
First, we know that $\emph{Q}$ is discrete from theorem \ref{th: Discrete in Cartesian product} and well-founded by lemma \ref{lem: WQO in Cartesian product}.
$D_i$ are well-ordered from which we can deduce that Q possesses a least element denoted by $l$ so we have $RS(M,Init) \subseteq [l, r]$.
Since $D_i$ are well-ordered, we can deduce that each projection of $[l, r]$ on $D_i$ is a closed interval $[l_i, r_i]$, hence is compact by corollary \ref{cor: Compactness+WO}. 
Then, by Tychonoff's theorem \ref{th: Tychonoff}, we deduce that $[l, r]$ is also compact as a Cartesian product of $d$ compacts. 
Being also discrete as a subset of $\emph{Q}$, we can conclude that $[l, r]$ is an interval both discrete and compact, therefore, it is finite by theorem \ref{th: Compact+discrete = finite}. 
Since  $RS(M,Init) \subseteq [l, r]$, $RS(M,Init)$ is also finite
\hfill
$\square$

Note that if $\mathcal{SV}$ is infinite, then we cannot conclude that $\emph{Q}$ is discrete. Actually, counterexamples does exist in the general case.

This theorem does apply to a large number of similar models such as Petri nets (since Petri nets are SVTSs with a finite number $d$ of state variables (i.e. of places)), VAS, Vector Replacement Systems \cite{Kel72,Mayr81}, or P/T systems \cite{Reisig87} to name a few.
They only differ by constraints put on the transition relation and their states, being defined as elements of $\mathbb{N}^d$, straightforwardly satisfy the constraint of discreteness and well-ordering of theorem \ref{th: finiteness=boundedness in SVTS}.


\section{Conclusion}
\label{sec: conclusion}
To establish a significant space where finiteness and boundedness coincide, we had to turn to topology theory and prove theoretical results that are new to the best of our knowledge, in particular that if a multi-dimensional interval $[\mathbf{a}, \mathbf{b}]$ inside a finite product of well-ordered sets is discrete, collapses directly into a finite set. 
This formally bridges the gap between order-theoretic discrete structures and general topological constraints.

Then, we introduced a new class of transition systems that include state variables and is directly inspired from the works of A. Pnueli and Z. Manna with the aim of defining as broad as possible a class of transition systems where finiteness and boundedness coincide. Indeed, this provide us a two prong approach to solve the boundedness problem.
We have seen that this is generally not the case for WSTS
and that the monotonicity rule (which is essential to the definition of the WSTS and serves as the cornerstone of many results on decidability) does not apply in the context of this result. Hence, the introduction of VSTS without specific property for $\rightarrow$.

In regard to VSTS, it remains to better position this new conceptual machine with the rest of the vast literature on DES.

\bibliography{biblio}  
\end{document}